\documentclass[]{revtex4}

\usepackage[T1]{fontenc}
\usepackage{isolatin1}
\usepackage{graphicx}
\DeclareGraphicsExtensions{.pdf,.mps,.png,.jpg,.gif,.eps} 

\newcommand{\bfm}[1]{\mbox{\boldmath$#1$}}    
\newcommand{\bfmcal}[1]{\mbox{\boldmath$\cal #1$}}       

\begin{document}


\title{Comment on :  ``Neutrino Velocity Anomalies: A
Resolution without a Revolution''}

\author{Denis Bernard}
\affiliation{Laboratoire Leprince-Ringuet, Ecole Polytechnique, CNRS/IN2P3, F-91128 Palaiseau, France }
 
\date{\today}%

\begin{abstract}
I comment on a recent preprint ``Neutrino Velocity Anomalies: A
Resolution without a Revolution''
 that appeared recently as arXiv:1110.0989 [hep-ph]

\end{abstract}


\maketitle

\section{Introduction}

After the surprising result posted recently by the OPERA collaboration 
\cite{:2011zb}, 
Naumov and 
Naumov have posted an interpretation after which 
``{\em the neutrino advance of time observed in MINOS and OPERA experiments
can be explained in the framework of the standard relativistic quantum
theory as a manifestation of the large effective transverse size of
the eigenmass neutrino wavepackets}''  \cite{Naumov:2011mm}.

In that interpretation the wave packet of the traveling neutrino is
described as the product of a transverse and of a longitudinal
probability function \cite{Naumov:2011mm},
 without any curvature, something which is certainly true at the
location of the production (i.e., the source at CERN), but something
that might be questioned after propagation has taken place.

One consequence of the approximation made in \cite{Naumov:2011mm} is
that one side of the wave packet of a neutrino that would be traveling
towards the target detector with a small angle, would be propagating
with a superluminal velocity (see Fig. 1 of \cite{Naumov:2011mm}).

In this note I try and revisit the approximation made in
 \cite{Naumov:2011mm}  and its consequences.

\section{Packet wave propagation : back to basics}

Let's  consider the paraxial propagation of a wave packet from CERN to
Grand Sasso.
For the sake of the present note, let's consider the neutrino as a
regular particle in the frame of the standard model, in particular
that it be very respectful of special relativity.
The propagation of the wave packet is then governed by the
Klein-Gordon equation.  
For a multi-GeV energy, sub-eV/$c^2$ mass particle I will allow myself
to neglect the mass, and eventually get to Maxwell equation :
\begin{equation}
\nabla^{2}{\bfmcal E}({\bfm r},t)
-\frac{1}{c^{2}}\frac{\partial^{2}{\bfmcal E}({\bfm r},t)}{\partial t^{2}} = 0
\end{equation}

I will then follow the derivation of paraxial propagation, as can be
found in any textbook on lasers physics such as Ref.  \cite{Laser}.
Spin, and therefore polarization, are neglected here.
Separating the time and space variations \cite{Laser}, 
${\bfmcal E}({\bfm r},t) = {\bfmcal E}({\bfm r}) e^{i\omega t}$, 
we obtain the Helmholtz equation :
\begin{equation}
\nabla^{2}{\bfmcal E}({\bfm r}) + k^{2}{\bfmcal E}({\bfm r}) = 0,
\end{equation}
with $\omega = kc$.
In the paraxial approximation, that is the subject of this discussion, 
  solutions are searched under the parametrization of the form 
${\bfmcal E}({\bfm r})=  U(\rho,z)\exp{(-ikz)}$,
where $\rho \equiv \sqrt{x^2+y^2}$ describes the polar variable in the transverse plane
$(Oxy)$, and  $U(\rho,z)$
is a function that described the transverse distribution of the field
as a function of $z$. 
We have factorized the term $\exp{(-ikz)}$, which is the 1rst order
 phase term of something that propagates along $z$ with wavenumber
 $k$.
Assuming a slow variation of the envelope of the intensity along the axis, 
we obtain the paraxial wave equation  \cite{Laser}:
\begin{equation}
\left(\frac{\partial^{2}}{\partial \rho^{2}}+ \frac{1}{\rho}
\frac{\partial}{\partial \rho}\right)U- 2 i k\frac{\partial U}{\partial z}= 0
\end{equation}

The solutions to this equation that have a Gaussian profile $U \propto
\exp{(-\rho^{2}/w^{2})}$ are \cite{Laser}:
\begin{equation}
{\bfmcal E}(\rho,z)= {\cal E}_{0}
\frac{w_{0}}{w}\exp{(-\rho^{2}/w^{2})} \exp{\left[-i\left((kz -\phi) +
\frac{k\rho^{2}}{2R}\right)\right]} ,
\end{equation}
with : 
\begin{equation}
\phi= \arctan{\left(\frac{\lambda z}{\pi w_{0}^{2}}\right)} 
    = \arctan{\left(\frac{z}{z_{0}}\right)} ,
\end{equation}
\begin{equation}
R = z \left[1+\left(\frac{\pi w_{0}^{2}}{\lambda z}\right)^{2}\right]
  = z+ z_{0}^{2}/z, \label{eqray}
\end{equation}
and~:
\begin{equation}
w=w_{0}\left[1+\left(\frac{\lambda z}{\pi w_{0}^{2}}\right)^{2}\right]^{1/2}
 =w_{0}\left[1+\left(\frac{z}{z_{0}}\right)^{2}\right]^{1/2}
\end{equation}

$w(z)$ describes the transverse size of the beam.
On axis, the field amplitude varies like $w_{0}/w$, that is like
${\displaystyle 1/\sqrt{1+\left(\frac{\lambda z}{\pi
w_{0}^{2}}\right)^{2}}}$.
The intensity is a  Lorentzian function of $z$ with HWHM :
\begin{equation}
z_{0}= \pi w_{0}^{2}/\lambda , 
\end{equation}
named the Rayleigh length, or  the betatron
function at the production location.
Far from the source, 
 $w \sim \lambda z/\pi w_{0}$, which allow to define an angular divergence 
$w'=w/z \sim \lambda/\pi w_{0}= w_{0}/z_{0}$.  

~

The phase term $-i (kz-\phi+k\rho^{2}/2R)$ contains three
contributions :
\begin{itemize}   
\item $kz$ describes forward propagation (together with $-\omega t$)
along $z$.
\item  $-\phi$
 describes a phase variation with $z$
that induces a phase jump of $\pi$ through the wavepacket waist.
\item the last term $k\rho^{2}/2R$ with $R = z+ z_{0}^{2}/z$ is the
term of interest.
\end{itemize}   

Let's examine them further;
\begin{itemize}   
\item 
Close to the source ($z\approx 0$), the third contribution is
 $k\rho^{2}z/2 z_{0}^{2}$, of the order of $z/z_{0}$, that is extremely
 small.
The phase surfaces are therefore asymptotically planes perpendicular to $z$.
  
\item 
Far from the source, 
 ($z \rightarrow \infty$),
neglecting the 2nd, constant, contribution, we get a phase that goes like 
 $k(z+ \rho^{2}/2R)$.

For a phase surface, $\varphi=\varphi_{0}$, referenced by the phase on
axis $\varphi_{0}= \varphi(\rho = 0)$, the sphere with radius $R$ 
is described close to the axis by 
$(z-z_{a}) = -(1- \cos{\theta})R=-\theta^{2}R/2$ with $\rho=\theta R$,
that is $(z-z_{a}) = - \rho^{2}/2R$ and $(z_{a}=\varphi_{0}/k)$. 
The sphere is therefore described by $z + \rho^{2}/2R=\text{Cste}$.

Phase surfaces are therefore spheres with radius $R$, centered on the
source, with $R$ increasing linearly with $z$.
\end{itemize}   

Far from the source, the wavepacket is therefore described by a
paraxial spherical wave with a Gaussian transverse distribution.

\section{Discussion}

The wakepacket beam being asymptotically a plane wave right after
 production on a longitudinal range of the order of the Rayleigh
 length $z_0$, and a spherical wave asymptotically for $z \gg z_0$,
 let's estimate the numerical value of $z_0$.
\begin{itemize}   
\item 
For an Heisenberg-limited wavepacket, $z_0$ could be estimated to
  $\lambda  /(\pi w'^2)$, where the de Broglie length $\lambda $ is of
  the order of $10^{-2}$ fm for a 10 GeV beam,
and the divergence of the order of (1 km / 730 km)\cite{Naumov:2011mm},
that is $10^{-3}$.
Tiny.
\item 
For a non-Heisenberg-limited wavepacket, $z_0$ is obtained simply by
the ratio of the size of the source (centimeters) to the asymptotic
divergence ($10^{-3}$)\cite{Naumov:2011mm}, that is, tens of meters.
\end{itemize}   

The waist region close to the source, where the wavepacket beam is
asymptotically plane, is very small compared to the actual flight
length.

\section{Conclusion}

Using the usual treatment of the propagation of a wavepacket, as
developed for example in the frame of laser physics, we obtain an
asymptotic description as a spherical wave with a Gaussian transverse
profile, as soon as it has left the region close to source.

The quantum mechanics isochronous surface is therefore asymptotically
the same as the isochronous surface in classical mechanics, and the
propagation time simply determined by the distance from the production
point to the detection point, and the velocity of the traveling thing.

In Ref. \cite{Naumov:2011mm}, no attempt to compute the curvature of
the wavepacket was performed, and therefore no curvature was found,
which lead unavoidably to the claim of a superluminal effect.

\end{document}